# New Perspectives and Systematic Approaches for Analyzing Negative Damping-Induced Sustained Oscillation

Chongbin Zhao, *Student Member*, and Qirong Jiang

*Abstract*—Sustained oscillations (SOs) are commonly observed in systems dominated by converters. Under specific conditions, even though the origin of SOs can be identified through negative damping modes using conventional linear analysis, utilizing the describing function to compute harmonic amplitude and frequency remains incomplete. This is because a) it can not cover the cases where hard limits are not triggered, and b) it can not provide a complete trajectory for authentic linear analysis to confirm the presence of SO. Hence, two analytical methods are proposed by returning to the essential principle of harmonic balance. a) A dedicated approach is proposed to solving steady-state harmonics via Newton-Raphson iteration with carefully chosen initial values. The method encompasses all potential hard limit triggered cases. b) By employing extended multiharmonic linearization theory and considering loop impedance, an authentic linear analysis of SO is conducted. The analysis indicates that the initial negative damping modes transform into multiple positive damping modes as SO develops. Simulation validations are performed on a two-level voltage source converter using both PSCAD and RT-LAB. Additionally, valuable insights into the work are addressed considering the modularity and scalability of the proposed methods.

*Index Terms*—Sustained oscillation, Newton-Raphson iteration, multiharmonic linearization, loop impedance, harmonic balance, stability analysis.

## I. INTRODUCTION

Conventional linear analysis has been widely applied to study the very common issue of harmonic stability with the rising penetration of power electronics converters [1]-[11]. The system is initially assumed to be stable without any harmonic at the point of common coupling (PCC). If the negative damping mode is identified, the system will mostly diverge to a final periodic/steady state of *sustained oscillation* (SO). The authors in [4] reproduced the sub-super synchronous interactions between the type-IV wind turbine generation and the weak grid that have been widely reported in northwestern China since 2014. In Fig. 1 (b), an evident divergence is observed, and (negative damping-induced) SO appears when the d-axis inner loop control signal clips the upper hard limit. A similar phenomenon was also observed between a Static Synchronous Compensator (STATCOM) and the ac/dc grid in

China Southern Grid [5]. As Fig. 2 (b) shows, when the STATCOM was put into operation, the q-axis inner loop control signal rapidly reached the lower hard limit, and then a convergence with positive damping emerged before the system reached SO. Considering the serious hazard to relay protection and the practical design for self-adaptive oscillation mitigations [12], [13], the theoretical analysis to gain more insights into SO has attracted many research interests.

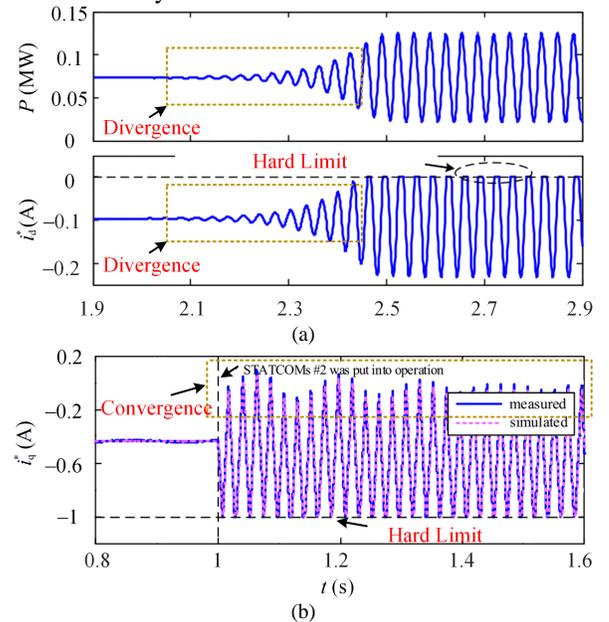

Fig. 1. Two real-world SOs. (a) From [4]. (b) From [5].

A preference for SO analysis is using the conclusions of conventional linear analysis [4], [14]. However, such a process can only obtain a satisfactory estimation of the initial point of oscillation. When the increasing harmonics are injected into both the power and control stages of the converter, the system dynamic response dynamically changes due to multiple nonlinearities [15], [16], which causes a deviation of the final steady state from the initial steady state. Hence, both the amplitude and frequency of SO cannot be effectively solved via linear analysis, and new analytical methods are required.

Currently, the describing function (DF) serves as the major theoretical basis for solving the steady-state harmonics of SO [17]-[22]. Inspired by Fig. 1 that a hard limit is often triggered

This work was supported in part by the National Natural Science Foundation of China under Grant U22B6008. (*Corresponding author: Chongbin Zhao*).

The authors are with the Department of Electrical Engineering, Tsinghua University, Beijing 100084, China (e-mail: zhaocb@mails.tsinghua.edu.cn; qrjiang@mail.tsinghua.edu.cn).

in the SO, the DF is used to describe the hard nonlinearity and forms a forward path, while the remaining control and power stages form an approximate linear feedback path, then the generalized Nyquist criterion is applied to realize graphical prediction of SO. The DF can also be integrated into the large-signal impedance model [22], [23] and ensures the solvability of the steady-state harmonic using iterations. Even if the DF is supported by the classical control theory, it is pointed out that the DF-based SO analysis is *incomplete* because:

a) The trigger of a hard limit is an insufficient condition of SO as demonstrated in [14], [15], [24] or indicated in other types of SO [3], [25], [26]. For the hard limit non-triggered case, the DF-based method is ineffective and no alternative has been reported. From another perspective, for the DF-based method, determining the triggered hard limit requires simulation and deviates from the objective of theoretical analysis. Similar deficiencies also exist in qualitative SO analyses using bifurcation theory [24], [27].

b) There are several unproven assumptions in the DF-based method. For example, the effects of soft nonlinearities such as trigonometric functions used in the phase-locked loop (PLL) are fully neglected, but the PLL plays an important role in the sub-super synchronous oscillation [4], [5]. Moreover, the DF requires a low-pass characteristic of the feedback path to avoid the influence of high-order SO components, but the high-order SO components can truly be observed [16], [21].

c) There is a similarity between the layout of the loop gain model [11] (a transfer function for conventional linear analysis) and the DF model. One may mix the two subjects and readily believe any calculated SO truly exists. To the author's knowledge, the critical step of confirming the existence of SO through an authentic linear analysis is lacking. In [15], SO is explained as the system being critically stable, but the explanation is dubious since any critically stable system cannot be truly observed. An intrinsic reason for the deficiency is that complete operating trajectory cannot be solved since only one closed loop is formed by a DF, and only the input/output of the triggered hard limit can be obtained. It also indicates the lack of flexibility of the DF-based method in dealing with the asymmetrical hard limits triggered case.

After reviewing the state-of-art, one should realize the lack of proper perspectives and systematic approaches for the SO analysis. This paper aims to fill the following two voids:

a) Regarding the steady-state calculation of SO, a set of nonlinear equations is established based on the very fundamental *harmonic balance*, and the variables are solved via Newton-Raphson iteration. The hard limit non-, unilateral, and bilateral triggered cases are fully covered. The selection of variables, handling of nonlinearities, and determination of initial values of iteration are introduced in detail.

b) Regarding the linear analysis of SO, based on the obtained steady-state harmonics, the multiharmonic linearization is extended to the system involving both fundamental and arbitrary (inter)harmonics. The frequency responses of a closed-loop impedance are obtained to identify the positive damping mode of the real SO.

The schematic diagram of Fig. 2 clearly explains the structure and contributions of this work, and the rest of this paper is organized as follows. Section II states the critical observation from simulated SO examples. Section III focuses on an accurate steady-state calculation of SO with multi-type nonlinearities considered and minimal simplifications. Section IV focuses on the authentic linear analysis of SO with high modularity. The proposed theories and methods are validated using simulations in Section V. Discussions and conclusions are addressed in Section VI.

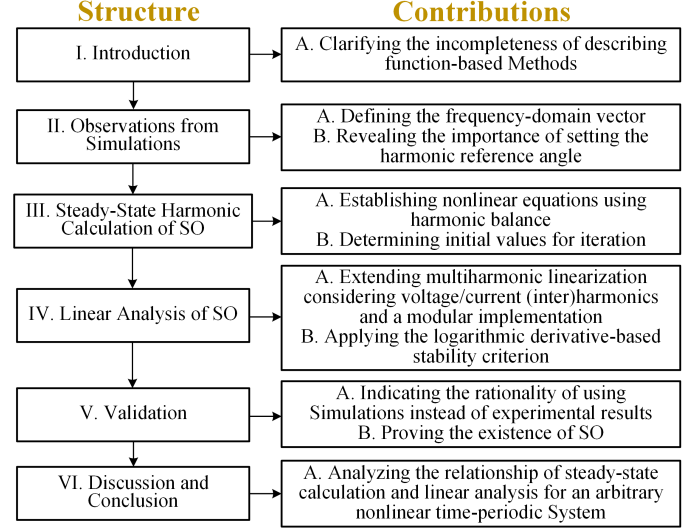

| Structure | Contributions |
|---|---|
| I. Introduction | A. Clarifying the incompleteness of describing function-based Methods |
| II. Observations from Simulations | A. Defining the frequency-domain vector B. Revealing the importance of setting the harmonic reference angle |
| III. Steady-State Harmonic Calculation of SO | A. Establishing nonlinear equations using harmonic balance B. Determining initial values for iteration |
| IV. Linear Analysis of SO | A. Extending multiharmonic linearization considering voltage/current (inter)harmonics and a modular implementation B. Applying the logarithmic derivative-based stability criterion |
| V. Validation | A. Indicating the rationality of using Simulations instead of experimental results B. Proving the existence of SO |
| VI. Discussion and Conclusion | A. Analyzing the relationship of steady-state calculation and linear analysis for an arbitrary nonlinear time-periodic System |

Fig. 2. Schematic diagram of the work.

## II. OBSERVATIONS FROM SIMULATIONS

### A. System Overview

The basic scenario of a two-level voltage source converter (TL-VSC) fed by a three-phase symmetric grid is studied. The setup and control loops are shown in Fig. 3, while the parameters are listed in Table I. The conventional linear analysis for the system has been thoroughly discussed in [7], which proves that a pair of negative damping modes emerges when $L_g$>1 mH. To fully reveal the effectiveness of the proposed SO analysis, four tests are created by changing $L_g$ cooperating with the limit values $i_{up}/i_{low}$ in Table I.

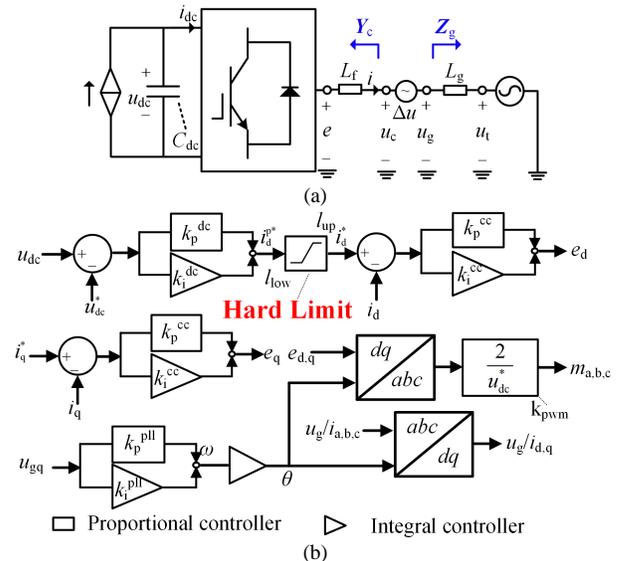

Fig. 3. Testing system. (a) Setup. (b) The studied control loop.

TABLE I
PARAMETERS OF THE TESTING SYSTEM

| Symbols | Meanings | Test 1 | Test 2 | Test 3 | Test 4 |
|---|---|---|---|---|---|
| $L_g$ | grid inductor | 1 mH | | | 1.5 mH |
| $i_{up}$ | upper limit of $i_a^*$ | 500 | 200 | 200 | 200 |
| $i_{low}$ | lower limit of $i_a^*$ | −500 | −500 | −500 | 0 |
| $U_t$, $I_{load}$ | ac voltage, dc current source | 380 V, 66.66 A | | | |
| $L_f$, $C_{dc}$ | filter inductor, dc capacitor | 0.55 mH, 5 mF | | | |
| $u_{dc}^*$, $i_q^*$ | dc voltage/q-axis current reference | 750 V, 0 A | | | |
| $f_1$ | nominal frequency | 50 Hz | | | |
| $k_p^{dc}+k_i^{dc}s^{-1}$ | dc voltage PI controller | 0.1+100$s^{-1}$ | | | |
| $k_p^{ac}+k_i^{ac}s^{-1}$ | ac current PI controller | 0.1+10$s^{-1}$ | | | |
| $k_p^{pll}+k_i^{pll}s^{-1}$ | PI controller of PLL | 3+100$s^{-1}$ | | | |

## B. Simulations

Simulations performed in PSCAD can describe the reported features and neglected details [15], [16], [24] of SO. As Fig. 4 (a) shows, the system diverges when $L_g$ is switched from 0.1 to 1 mH at $t$=2 s. The observations are as follows:

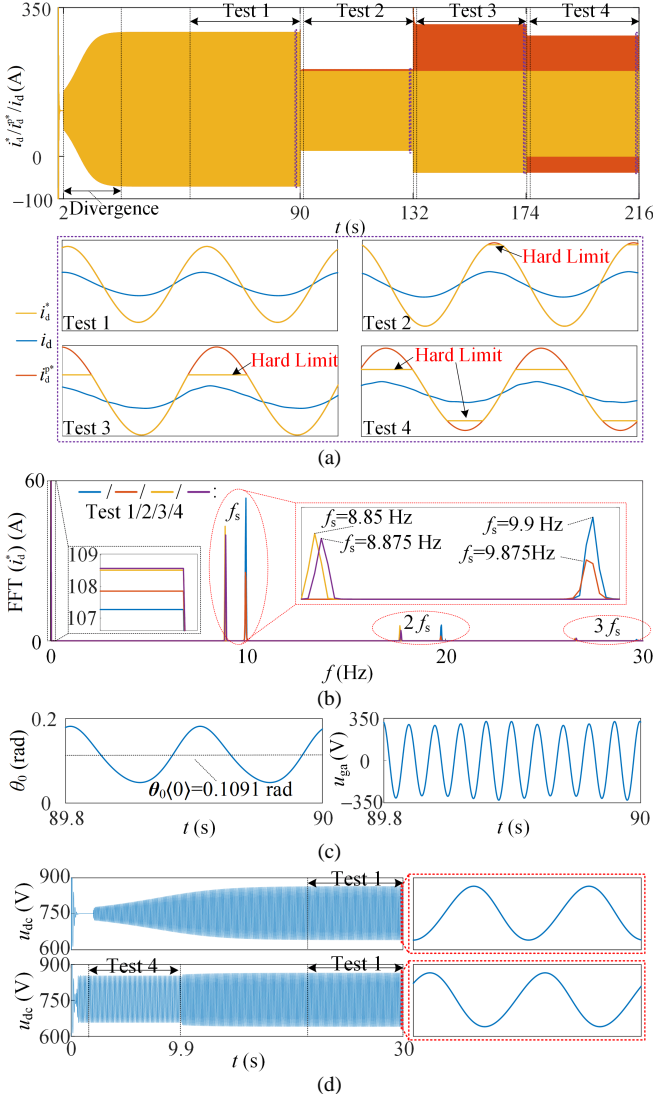

Fig. 4. Simulation results. (a) Comparison between $i_a^*$, $i_a$, and $i_a^{*'}$. (b) Harmonic distribution of dc steady-state signal. (c) Waveforms of $\theta_0(t)$ and $u_{ga}(t)$. (d) The waveform of $u_{dc}(t)$ under various settings.

**a) Distributions of the generalized ac/dc steady-state signals**: For SO, a generalized ac (dc) signal is defined as a time-domain signal with a (non)-zero dc component. Suppose that a dc steady-state signal contains the dominant 1st-order harmonic at $f_s$, it also contains the 0th-order (0), 2nd-order (2$f_s$), 3rd-order (3$f_s$), ···, harmonics. Correspondingly, a generalized ac steady-state signal contains the 0th-order ($f_1$), 1st-order ($f_1\pm f_s$), 2nd-order ($f_1\pm 2f_s$), 3rd-order ($f_1\pm 3f_s$), ··· harmonics. There are theoretically infinite harmonics but the amplitudes of higher-order components decrease due to the coupling mechanism [8], so a manmade Nth-order truncation is used for each time-domain signal $g(t)$ for the calculation. $g(t)$ can be transformed into a frequency-domain vector $\mathbf{g}$ using the Fourier coefficients $\mathbf{g}\langle\cdot\rangle$:

$$\mathbf{g} = [\mathbf{g}_{-1}, \mathbf{g}_0, \mathbf{g}_1]_{[3\times(2\times2N+1)d]^T},$$
$$\mathbf{g}_k = [\mathbf{0}_{N\times1}, \mathbf{g}\langle k-Ns \rangle, \mathbf{g}\langle k-(N-1)s \rangle, \cdots, \mathbf{g}\langle k-2s \rangle, \mathbf{g}\langle k-s \rangle, \mathbf{g}\langle k \rangle, \quad (1)$$
$$\mathbf{g}\langle k+s \rangle, \mathbf{g}\langle k+2s \rangle, \cdots, \mathbf{g}\langle k+(N-1)s \rangle, \mathbf{g}\langle k+Ns \rangle, \mathbf{0}_{N\times1}]_{[2\times2N+1)d]^T}.$$

where k=−1, 0, 1. k and Ns are used to express the frequency-shift property: k represents k times of $f_1$ while Ns represents N times of $f_s$. $\mathbf{0}$ is the zero vector to prevent the overflow due to the multiplication in the matrix operation.

A generalized ac/dc signal contains the positive--negative-/zero-sequence (PS-NS/ZS) components at an arbitrary frequency. Hence, $\mathbf{g}_1/\mathbf{g}_0/\mathbf{g}_{-1}$ in (1) is defined as the generalized PS/ZS/NS steady-state vector. $\mathbf{g}$ is conjugate symmetric; $\mathbf{g}_{\pm1}/\mathbf{g}_0$ is a zero vector for the generalized PS-NS/ZS steady-state vector; using the one-phase generalized PS vector can represent the three-phase ac signal in the follow-up operations [9]. In this work, the generalized ZS steady-state vectors include $i_{d,q}$, $i_{d,q}^{p^*}$, $i_{load}$, $e_{d,q}$, $u_{dc}$, $u_{gd,q}$, and $\theta_0$, while the generalized PS/NS steady-state vectors include $i_a$, $u_{ta}$, $u_{ga}$, $u_{ca}$, $m_a$, $\cos(\theta)$, and $\sin(\theta)$.

**b) Deviations of $f_s$ and 0th-order harmonics for SO from the initial state**: The estimated $f_s$s of the initial states ($f_{s0}$s) are 9.9132 ($L_g$=1 mH) and 9.8366 ($L_g$=1.5 mH) Hz [7]. $f_{s0}$s deviate from the results of Fast Fourier Transforms (FFTs), especially for the hard limit triggered case in Tests 2-4, as Figs. 4 (a) and (b) show. Deviations also exist in 0th-order signals (such as $i_a^*$ in Fig. 4 (b) and $\theta_0(0)$ in Fig. 4 (c) ($\theta_0(0)$ deviates from 0.1097 rad of the initial state)). Hence, $f_s$ and 0th-order signals should be regarded as unknown variables for the steady-state SO calculations.

**c) Phase shift of high-order harmonics**: In Fig. 4 (d), an extra simulation is presented and Test 4 is switched to Test 1 at $t$=9.9 s. Comparing the $u_{dc}(t)$ from 29.8 to 30.0 s in the two simulations, the amplitude of each order harmonic is consistent, but an evident phase shift exists only for high-order harmonics. This is because the ideal ac grid only contains the fundamental harmonic and has set a fundamental reference angle (assumed as 0° for the steady-state calculation). It also indicates that a separate harmonic reference angle should be set for the steady-state calculation of SO. Such an indication is reasonable since only the oscillation amplitude and frequency instead of the specific phase is of interest for the SO analysis.

## III. Steady-State Harmonic Calculation of SO

### A. Establishment of Nonlinear Equations

The essence of solving steady-state harmonics of SO is to solve the Fourier coefficients based on (1). Each coefficient introduces two variables and can be expressed in either polar or rectangular coordinates as $M(\cdot)\angle A(\cdot)$ or $R(\cdot)+jI(\cdot)$. Table II lists the targeted vectors that prove the minimum number and derivability over the equilibrium of nonlinear equations.

The multiplication of time-domain signals is equivalent to the convolution of frequency-domain vectors, and the latter can be transformed into the multiplication of the Toeplitz matrix $\boldsymbol{T}(\cdot)$ of one vector with the other vectors [8]. Hence, the nonlinear equations can be extracted from the frequency-domain operations by separating the real and imaginary parts.

The approximations of each type of nonlinearity are worth addressing:

**a) Manmade truncation for the infinite product coupling of modulation**: Similar to the conventional impedance modeling of modular multilevel converter [8], [10].

**b) Approximating trigonometric functions with input harmonics for Park transformation**: The details are offered in Appendix A, part a). A separate order determination for this nonlinearity is emphasized. Due to the form of the Bessel function, the maximum harmonic order of $\cos[\theta(t)]/\sin[\theta(t)]$ increases at a square rate as N increases but the harmonic orders of the other signals increase at a linear rate as N increases. Even if considering the setting zero vector in (1), when setting the maximum N=3, N of $\theta_0(t)$ is recommended to be less than 2 regardless of N of other signals.

**c) Input-output modeling of triggered hard limit (optional)**: The details are offered in Appendix A, part b). Supposing a continuous input signal, the triggered hard limit leads the output signal to be piecewise. A major assumption is to neglect the high-order harmonics and use the dominant 1st-order harmonic to calculate the segment moments [21]. The assumption will not bring evident error considering the typical small amplitude of high-order harmonics. It can also yield a derivable expression for the output signal.

The details of intermediate vectors and extra equations for constructing the nonlinear equations are offered in Appendix A, part c). A numerical example of the hard limit triggered case with N=3 is focused on Table II. The table explicates the principle of establishing the nonlinear equations and extracting the targeted variables. The table also clarifies the feasibility of the solutions are nonzero.

applying Newton-Raphson iteration since the total number of nonlinear equations and targeted variables are equal (both 45). Please note that the process of forming nonlinear equations in Table II majorly corresponds to the three types of nonlinearities (product coupling of modulation, for Groups I & II; Park transformation, for Groups III & IV; triggered hard limit, for Group V).

### B. Determination of Initial Values for Iteration

Systematically determining the initial values is a key point of Newton-Raphson iteration, and a feasible approach is proposed for the hard limit non-triggered case:

a) Set n=0 and calculate the steady-state harmonic for conventional linear analysis, which also requires a simple Newton-Raphson iteration [7]. If a group of negative damping modes is identified (40 & 60 Hz at ac side [7]), use the transfer immittance [6] to identify the dominant physical quantity of oscillation. The transfer immittance $\boldsymbol{Y}_{TI}$ of the TL-VSC under dc voltage control mode is [7]:

$$\begin{bmatrix} \Delta i_n(s) \\ \Delta i_p(s) \\ \Delta u_{dc}(s) \end{bmatrix} = \boldsymbol{Y}_{TI}^{(3\times3)} \begin{bmatrix} \Delta u_n(s) \\ \Delta u_p(s) \\ \Delta i_{dc}(s) \end{bmatrix}, \begin{bmatrix} \Delta u_n(s) \\ \Delta u_p(s) \\ \Delta i_{dc}(s) \end{bmatrix} = \boldsymbol{Z}_{TI}^{(3\times3)} \begin{bmatrix} \Delta i_n(s) \\ \Delta i_p(s) \\ \Delta u_{dc}(s) \end{bmatrix}, \boldsymbol{Y}_{TI} = \boldsymbol{Z}_{TI}^{-1}. \quad (2)$$

where the prefix $\Delta$ represents a small-signal perturbation while subscript p/n refers to the PS/NS. As Fig. 5 shows, the magnitude of $\boldsymbol{Y}_{TI}(3,1)/\boldsymbol{Y}_{TI}(3,2)$ is greater than that of $\boldsymbol{Z}_{TI}(1,3)/\boldsymbol{Z}_{TI}(2,3)$ below 100 Hz. Neglect the currents with small amplitude and use the superposition theorem, the 1 p.u. ac voltage perturbation will lead to a dc voltage perturbation of over 1 p.u., so it is predicted that the dc voltage oscillation is more prominent than the ac voltage oscillation at the beginning of negative damping oscillation. Furthermore, regarding searching the acceptable set of initial values for Newton-Raphson iteration, scan $\boldsymbol{u}_{dc}(s)$ can achieve the balance between a fast speed and a great possibility.

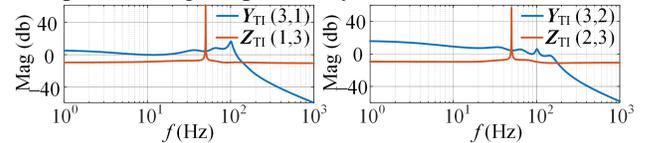

Fig. 5. Coupling terms of the transfer immittance.

b) Set n=1 and use the results of n=0 as the initial value of each 0th-order harmonic and $f_s$; scan $R(\boldsymbol{u}_{dc}(s))$ and $M(\theta_0(s))$ with a step of 1% of $R(\boldsymbol{u}_{dc}(0))$ and $M(\theta_0(0))$ and set the initial values of the other 1st-order harmonics to 0. Conduct the iteration until the solutions are nonzero.

TABLE II
Principles of Applying Newton-Raphson Iteration (N=3, Hard Limit Triggered Case (Tests II-IV))

| Group | Variables | | | Equations | | |
|---|---|---|---|---|---|---|
| | Vector Number | | Comments | Relationship Number | | Comments |
| I | $\boldsymbol{m}_a$ | 14 | In rectangular coordinate; solving the generalized PS vector | $i_{dc}=i_{dc}'$ | 7 | Extracting from the generalized ZS vector |
| II | $\boldsymbol{u}_{dc}$ | 6 | In rectangular coordinate; solving the generalized ZS vector, $\boldsymbol{u}_{dc}(0)$ is known based on (A9-3) | $\boldsymbol{e}_a=\boldsymbol{e}_a'$ | 13 | Extracting from the generalized PS vector, 1 linear equation is removed based on (A10) |
| III | $\boldsymbol{u}_{ga}$ | 13 | In rectangular coordinate; solving the generalized PS vector, $I(\boldsymbol{u}_{ga}(1))$ is calculated based on (A10) | $\boldsymbol{e}_{d/q}=\boldsymbol{e}_{d/q}'$ | 12 | Extracting from the non-central rows of the generalized ZS vector based on (A9-1) |
| IV | $\theta_0$ | 4 | In polar coordinate; solving the generalized ZS vector, $A(\theta_0(3s))$ and $M(\theta_0(3s))$ are neglected based on Section III. A; $A(\theta_0(0s))$ is set to zero as the harmonic reference angle | $\theta=\theta'$ | 5 | Extracting from the generalized ZS vector based on Section III. A by separately setting N=2 |
| V | $\tilde{i}_d^r$ | 7 | In polar coordinate; solving the generalized ZS vector; should be blocked for the hard limit triggered case | $\tilde{i}_d^r=\tilde{i}_d^{r\prime}$ | 6 | Extracting from the generalized ZS vector based on (A9-3) |
| VI | $f/\omega$ | 1 | $f_s$ is the only unknown variable | $i_{d/q}(0)=i_{a/b}(0)$ | 2 | Based on (A9-1) to complement $\boldsymbol{e}_{d/q}=\boldsymbol{e}_{d/q}'$ |

c) Set n=2 and use the results of b) as the initial value of each $0^{th}$, $1^{st}$-order harmonics, and $f_s$; scan $M(\theta_0\langle 2s\rangle)$ with a step of 1% $M(\theta_0\langle s\rangle)$ and set the initial values of other $2^{nd}$-order harmonics to be 0. Conduct the iteration until the solutions are nonzero, and follow the similar rule until n=N.

### C. Implementation

Fig. 6 presents the flowchart of the steady-state harmonic computing method of SO. Annotations are added below:

a) When initially constructing the equations, the hard limit is supposed not to be triggered, i.e., (A6) is blocked and $\hat{\boldsymbol{i}}_d^* = \hat{\boldsymbol{i}}_d^{p*}$. If the following relationship satisfies for the first iteration:

$$M(\hat{\boldsymbol{i}}_d^{p*}\langle 0\rangle) + 2M(\hat{\boldsymbol{i}}_d^{p*}\langle s\rangle) > l_{up} \parallel M(\hat{\boldsymbol{i}}_d^{p*}\langle 0\rangle) - 2M(\hat{\boldsymbol{i}}_d^{p*}\langle s\rangle) < l_{low}, \quad (3)$$

the hard limit is identified as triggered, and (A6) should be considered to construct the new set of nonlinear equations.

b) Theoretically, the initial value of iteration should be separately determined for hard limit triggered cases. However, observing Fig. 4 (a), a hard limit non-triggered case can transit into a hard limit triggered case, so the results of the hard limit non-triggered case can serve as the initial value of iteration for the hard limit triggered cases. Hence, starting from hard limit non-triggered SO calculation is rational.

c) It is recommended to set a domain for the parameter scan in Section III. B, e.g., 0-40% for $M(\cdot)$ and –40-40% for $R(\cdot)$. If acceptable initial values cannot be found, the hard limit is mostly triggered for such a case since a large voltage dip collapses the system [27]. To avoid determining the initial value of $\hat{\boldsymbol{i}}_d^{p*}$ for a hard limit triggered case, the alternative is to slightly decrease $L_g$ and reconstruct the nonlinear equations until a proper initial value can be found in the domain.

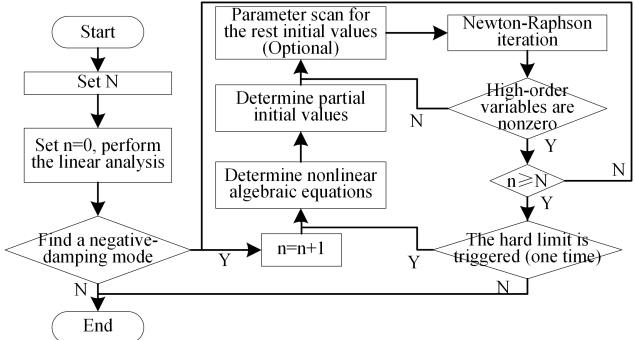

Fig. 6. Flowchart of the proposed steady-state harmonic calculation.

## IV. LINEAR ANALYSIS OF SO

### A. The Extended Multiharmonic Linearization Theory

Multiharmonic linearization was first applied to a modular multilevel converter considering the infinite coupling between the arm current and the cell capacitor [8], [10], and the same feature holds for SO. To extend the theory to the linear analysis of SO, the following concepts are clarified:

a) **Law of small-signal harmonic distribution**: *Suppose that a PS voltage perturbation at $f_p$ is added at the ac side*, the small-signal vector $\Delta\boldsymbol{g}$ is defined as:

$$\Delta\boldsymbol{g} = [\Delta\boldsymbol{g}_{p-2}, \Delta\boldsymbol{g}_{p-1}, \Delta\boldsymbol{g}_p]_{1\times(2\times2N+1)\times1}^T,$$
$$\Delta\boldsymbol{g}_{k'} = [\boldsymbol{0}_{N\times1}, \Delta\boldsymbol{g}\langle k'-Ns\rangle, \Delta\boldsymbol{g}\langle k'-(N-1)s\rangle, \cdots, \Delta\boldsymbol{g}\langle k'-2s\rangle, \Delta\boldsymbol{g}\langle k'-s\rangle, \Delta\boldsymbol{g}\langle k'\rangle, \quad (4)$$
$$\Delta\boldsymbol{g}\langle k'+s\rangle, \Delta\boldsymbol{g}\langle k'+2s\rangle, \cdots, \Delta\boldsymbol{g}\langle k'+(N-1)s\rangle, \Delta\boldsymbol{g}\langle k'+Ns\rangle, \boldsymbol{0}_{N\times1}]_{1\times2N+1)\times1}^T.$$

where $k'=p-2, p-1, p$. Similar to (1), $\Delta\boldsymbol{g}_{p-2}/\Delta\boldsymbol{g}_{p-1}/\Delta\boldsymbol{g}_p$ is defined as the generalized PS/ZS/NS small-signal vector. $k'$ indicates the frequency-shift property of $\Delta\boldsymbol{g}$ over $\boldsymbol{g}$. The difference between $k'$ and k reflects that $f_1$ in $\boldsymbol{g}$ is replaced by $f_p$ in $\Delta\boldsymbol{g}$, so a new vector $\boldsymbol{f}_p$ is defined:

$$\boldsymbol{f}_p = [f_p - 2f_1 + (-2N : 2N)f_s, f_p - f_1 + (-2N : 2N)f_s, f_p + (-2N : 2N)f_s] \quad (5)$$

b) **Law of matrices**: Toeplitz matrices are established based on the calculated steady-state harmonics. Diagonal transfer function matrices $\boldsymbol{Z}(s)/\boldsymbol{Y}(s)$ and $\boldsymbol{H}(s)$ are obtained by substituting $s$ with $j\omega_p=j2\pi\boldsymbol{f}_p$ into the transfer functions of passive elements and PI controllers. The two types of *full-rank matrices* are sufficient for multiharmonic linearization without any additional operation to block an element in the matrix [8].

### B. Loop Impedance Derivation

Loop impedance is a 1-dimensional closed-loop transfer function, and part of the system modes is embedded in the numerator polynomial of loop impedance. Either ac or dc loop impedance is fully observable to the complete modes of the testing system [7], so only the ac loop impedance $Z_{loop}(s)$ is modeled. Fig. 7 illustrates the relationship of $Z_{loop}(s)$ with the open-loop port impedance $\boldsymbol{Y}_c(s)$ and $\boldsymbol{Z}_g(s)$ as Fig. 3 (a) shows:

$$Y_{loop}(s) = -\frac{\Delta i(s)}{\Delta u(s)} = -\boldsymbol{A}\frac{\boldsymbol{Y}_c(s)}{\boldsymbol{I} + \boldsymbol{Y}_c(s)\boldsymbol{Z}_g(s)}\boldsymbol{B}, Z_{loop}(s) = Y_{loop}(s)^{-1}. \quad (6)$$

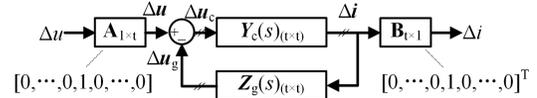

$$\boldsymbol{A}_{1\times1} = [0,\cdots,0,1,0,\cdots,0]$$
$$\boldsymbol{B}_{1\times1} = [0,\cdots,0,1,0,\cdots,0]^T$$

Fig. 7. Diagram of the loop impedance (t=3×(2×2N+1)).

The reasons for using $Z_{loop}(s)$ for SO linear analysis based on multiharmonic linearization include: a) $\boldsymbol{Y}_c(s)$ is infeasible to test since it is difficult to set the steady state of the converter under an ideal grid (with harmonics) and the SO to be the same; b) The multifrequency coupling can be fully considered; c) The modularity and scalability are high with the clear interpretation of each step.

To explicit the implementation, an example is offered to describe the effect of ac voltage dynamics $\Delta\boldsymbol{u}_{ca}$ on the insertion index dynamics $\Delta\boldsymbol{m}_a$ using the transfer function matrix $\boldsymbol{P}(s)$. Based on the fundamentals of impedance modeling [9], only PLL contributes to $\boldsymbol{P}(s)$ for the testing system in Fig. 3. The dynamics of PLL output $\Delta\theta$ is selected as an intermediate vector, and the frequency-domain linearization from $\Delta\boldsymbol{u}_{ca}$ to $\Delta\theta$ is given by:

$$\boldsymbol{H}_{pll}(s)\Delta\boldsymbol{u}_{cq} = \Delta\theta, \boldsymbol{u}_{cq} = -2\boldsymbol{T}(\sin(\theta^*))\boldsymbol{u}_{ca}$$
$$\Rightarrow \Delta\boldsymbol{u}_{cq} = -2\boldsymbol{T}(\sin(\theta^*))\Delta\boldsymbol{u}_{ca} - 2\boldsymbol{T}(\cos(\theta^*))\boldsymbol{u}_{ca}\Delta\theta$$
$$\Rightarrow \Delta\theta = -2\boldsymbol{H}_{pll}(s)\boldsymbol{T}(\sin(\theta^*))\Delta\boldsymbol{u}_{ca} - \boldsymbol{H}_{pll}(s)\boldsymbol{T}(\boldsymbol{u}_{ca})\Delta\boldsymbol{u}_{cq} \quad (7)$$
$$\Rightarrow \Delta\theta = \underbrace{-2[\boldsymbol{I} + \boldsymbol{H}_{pll}(s)\boldsymbol{T}(\boldsymbol{u}_{ca})]^{-1}\boldsymbol{H}_{pll}(s)\boldsymbol{T}(\sin(\theta^*))}_{G_{pll}(s)}\Delta\boldsymbol{u}_{ca}$$

where the superscript $''$ indicates that in the linear analysis, the generalized PS-NS steady-state vectors should be recalculated. The reason is that a shift of fundamental reference angle exists in the steady-state calculation and linear analysis as PLL requires. The function can be achieved by only forcing $A(\theta_0\langle 0\rangle)$ to 0 and performing the inverse Park transformation using the corresponding ZS steady-state vectors.

A separate validation is performed on $G_{\text{pll}}(s)_{(6N+2,\ 10N+3)}$: $f_s$ is set as 10 Hz; the magnitude of $1^{\text{st}}$- and $2^{\text{nd}}$-order harmonics of $u_c$ are 30% and 10% of that of the fundamental, respectively; please note there is an $f_1$ shift between the PS injections of $\Delta \boldsymbol{u}_c$ and the ZS responses of $\Delta \boldsymbol{\theta}$. In Fig. 8, by comparing the red circles measured using the frequency scan in PSCAD with the yellow solid line calculated by the theoretical frequency responses of (7) in MATLAB, the correctness of (7) is verified. Moreover, the discrepancies between the solid lines are obvious, especially for the phase responses, so it is necessary to consider the ac voltage harmonics in PLL modeling for the linear analysis of SO.

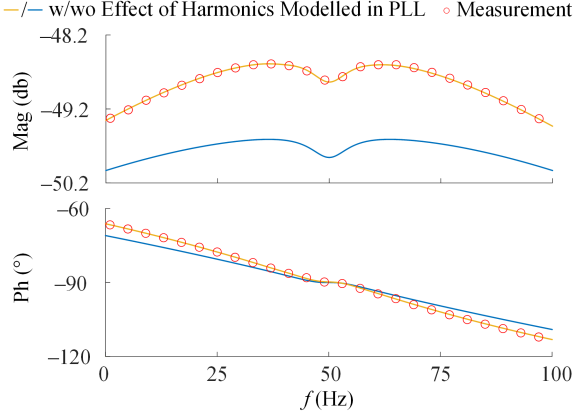

Fig. 8. Validation of $G_{\text{pll}}(s)_{(6N+2,\ 10N+3)}$.

The contribution of $\Delta \boldsymbol{\theta}$ to $\Delta \boldsymbol{m}_a$ related to the inverse Park transformation can be divided into two parts as (A8) indicates, one is related to $\boldsymbol{e}_{\text{dq}}$, and the other one is related to $\boldsymbol{i}_{\text{d,q}}$ combined with the PI control. Therefore, $\boldsymbol{P}(s)$ is expressed as:

$$\boldsymbol{P}(s) = -[\boldsymbol{T}(\cos(\boldsymbol{\theta}^r))][\boldsymbol{T}(\boldsymbol{e}_q) + \boldsymbol{H}_{\text{cc}}(s)\boldsymbol{T}(\boldsymbol{i}_q)] \\ + \boldsymbol{T}(\sin(\boldsymbol{\theta}^r))[\boldsymbol{T}(\boldsymbol{e}_d) + \boldsymbol{H}_{\text{cc}}(s)\boldsymbol{T}(\boldsymbol{i}_d)]\boldsymbol{G}_{\text{pll}}(s) \quad (8)$$

The above process can be extended to any other transfer function matrices in Appendix B. By performing the matrix operations at every $f_p$, the frequency responses of $Z_{\text{loop}}(s)$ are obtained by inverting that of the corresponding element in $Y_{\text{loop}}(s)$. Since matrix multiplication is not commutative, the matrix operation must follow the given equations. Moreover, the complete steady-state harmonic calculation in Section III serves as the solid basis of the linear analysis of SO.

### C. Mode Identification using Frequency Responses

Linear analysis is mostly used to confirm that no negative damping mode exists in the system, i.e., the calculated SO truly exists. The proposed criterion based on the *logarithmic derivative* of frequency responses [7] is briefly reviewed.

$Z_{\text{loop}}(s)$ can be written in the factored zero-pole form as:

$$Z_{\text{loop}}(s) = \frac{\prod_{n=1}^{N_z} a_{Zn}(j\omega - Z_n)}{\prod_{n=1}^{N_p} a_{Pn}(j\omega - P_n)} \quad (9)$$

where a, Z, and P (as well as the subscript) represent the flat gain, zeros, and poles, respectively.

The first-order numerator polynomial $(g_Z(\omega) = a_Z(j\omega - \lambda_Z)$, $\lambda_Z = \alpha_Z + j\omega_Z)$ is used to explain the basic concepts. Calculating the logarithmic derivative $(D_L(\cdot))$ of $g_Z(\omega)$ yields:

$$D_L(g_Z) = \frac{d[\log(g_Z)]}{d\omega} = \frac{d(g_Z)}{g_Z d\omega} = \frac{j}{j\omega - \lambda_Z} = \frac{j}{-\alpha_Z + j(\omega - \omega_Z)} \quad (10)$$

In (10), $a_Z$ is eliminated with only the information of the system mode left, and $D_L(\cdot)$ can be calculated using the difference method based on the third term as long as the frequency responses are available. Projecting the complex function to two real functions by separating the real and imaginary parts and calculating their $1^{\text{st}}$- and $2^{\text{nd}}$-order derivatives gives:

$$\text{Re}[D_L(g_Z)]\big|_{\omega=\omega_Z} = 0, \ \ \text{Im}[D_L(g_Z)]\big|_{\omega=\omega_Z} = -\alpha_Z^{-1},$$
$$\frac{d\{\text{Re}[D_L(g_Z)]\}}{d\omega}\bigg|_{\omega=\omega_Z} = \frac{1}{\alpha_Z^2}, \ \frac{d\{\text{Im}[D_L(g_Z)]\}}{d\omega}\bigg|_{\omega=\omega_Z} = 0, \quad (11)$$
$$\frac{d^2\{\text{Re}[D_L(g_Z)]\}}{d\omega^2}\bigg|_{\omega=\omega_Z} = 0, \ \frac{d^2\{\text{Im}[D_L(g_Z)]\}}{d\omega^2}\bigg|_{\omega=\omega_Z} = \frac{2}{\alpha_Z^3}.$$

Based on (11), the feature of $D_L(\cdot)$ is analyzed below:

a) If the positive (negative) damping mode exists in $g_Z$, i.e., $\alpha_Z < 0 \ (>0)$, a definite zero-crossing positive slope of $\text{Re}[D_L(g_Z)]$ and a maximum (minimum) of $\text{Im}[D_L(g_Z)]$ coexist at $\omega = \omega_Z$. Correspondingly, when $g_Z^{-1}$ is regarded as a basic unit of the denominator polynomial, all the results in (11) should multiply $(-1)$. Therefore, $\text{Re}[D_L(g_Z)]$ can be used to identify the existence of mode while $\text{Im}[D_L(g_Z)]$ can be used to confirm the positive/negative damping.

b) Thanks to the derivative operation and the quadric terms of $\omega$ in the denominator polynomial of $\text{Re}[D_L(g_Z)]$ & $\text{Im}[D_L(g_Z)]$ [7], $D_L(\cdot)$ transforms the multiplication & division to the addition & subtraction and realizes an approximate decoupling of each weak-damping zero and pole. Therefore, the damping can be reliably estimated using the peak values of $\text{Im}[D_L(Z_{\text{loop}})]$ by the second equation of (11). Distinguishing from the Nyquist criterion, the influence of the right-half plane pole on the stability analysis is excluded. Such a feature suits SO as Section IV. B discusses.

## V. Validation

### A. Considerations of Simulation Validation

For the SO analysis, it is believed that the simulation validation in various software (PSCAD and RT-LAB) is sufficient considering the following facts:

a) The pulse width modulation, the discrete digital control, or the uncertain natural passive damping is not considered in the modeling. The factors will not affect the proposed principle but may influence the experimental results as Table III indicates. In Table III, a clear simulation comparison between the converter switching model and the converter average model is performed in PSCAD. It is shown that only using a sufficient high switching frequency (over 20 kHz) can ensure a relative error below 2%. Such a phenomenon may distract one from the value of the work.

TABLE III
COMPARISON OF DOMINANT HARMONIC USING VARIOUS MODELS (TEST 1)

| Model | Switching | | | | | | Average |
|---|---|---|---|---|---|---|---|
| Frequency (kHz) | 1 | 5 | 10 | 20 | 50 | 100 | -- |
| $M(\boldsymbol{u}_{\text{dc}}(s))$ (V) | 46.4 | 50.5 | 53.8 | 55.5 | 55.9 | 56.0 | 56.0 |
| Relative Error (%) | 17.1 | 9.8 | 3.9 | 1.3 | 0.2 | ~0 | -- (Ref) |

b) The frequency scan using simulation benefits a more accurate measurement for SO. If one follows the common practice of setting the amplitude of voltage perturbation as 5% or 10 % to that of the fundamental voltage, the amplitude of voltage perturbation is comparable to that of the high-order harmonics. Based on trials and errors, it is recommended to set the voltage perturbation amplitude to 0.5% or 1 % of the fundamental voltage. It indicates that the unexpected noises will induce measurement errors in the prototype validation.

c) One can set a sufficient small simulation step (1 $\mu$s) to exclude the influence of numerical algorithms, and simulation results based on the converter average model given by the two platforms are very close. RT-LAB is specially used for frequency scan considering the faster simulation speed than PSCAD and the preference of decreasing the frequency resolution. Moreover, it is critical to avoid spectrum leakage for successful validation of SO by using the window function.

### B. Validation on Steady-State Harmonic Calculation

In Appendix C, the comparison between measurements and Newton-Raphson iteration (using the Symbolic Math Toolbox of MATLAB) for all four tests can be found.

Considering the importance of Test 1, the cases of N=0-3 are carefully studied using Table C1. "--" indicates that the corresponding variables are blocked in the nonlinear equations, and the number of equations increases as N increases. Key observations include:

a) The amplitudes of the high-order harmonics are very small. Setting the unknown harmonic order to N can generally ensure the accuracy of (N–1)[th]-order harmonics, so it is reasonable to set N=3 since the 2[nd]-order harmonics may increase in the hard limit triggered cases, as Fig. 4 indicates.

b) The FFT windows are set to be 40 s for each test so the resolution is 0.025 Hz, but the real $f_s$ cannot be probed for SO. Even if the window function is applied, the measurement is still an estimation of the real case, so the theoretical analysis is efficient and meaningful to reveal the essence of SO.

c) Table C1 also reflects the process of seeking initial values of iteration, as discussed in Section III. B. If $\mathbf{R}(\mathbf{u}_{dc}(\cdot))$ is not properly determined, the iteration converges to the result of N=0 and $f_s$ can be arbitrary. Such a case is unrealistic since the system diverges when N=0 [7]. It also indicates that the existence of any calculated SO should be separately verified using linear analysis even if obtaining nonzero harmonics.

In Tables C2-C4, only the results of N=3 are listed for each hard limit triggered case. As discussed in Section III. B, the results of Test 1 (N=3) can be selected as the initial value of iteration, because the obtained $\mathbf{i}_d^{o*}$ (=$\mathbf{i}_d^*$) in Test 1 triggers $l_{up}$ in Tests 2 &3 while both $l_{up}$ and $l_{low}$ in Test 4. With a larger $L_g$, the trigger should not change if the SO truly exists. 4-6 iterations are usually enough for accurate solutions, and the intermediate vectors can be obtained by substituting the results of Tables C1-C4 into (A6)-(A8). Such a step is necessary to confirm that none of the rest hard limits apart from the focused one is triggered. Moreover, $\mathbf{M}(\mathbf{i}_d^{o*}\langle s \rangle)$ must be monitored in the iteration to keep the inverse trigonometric functions solvable.

### C. Validation on Linear Analysis (Test 1)

The theoretical derivation of loop impedance based on the extended multiharmonic linearization with mode identification is mainly verified using Test 1, as shown in Fig. 9.

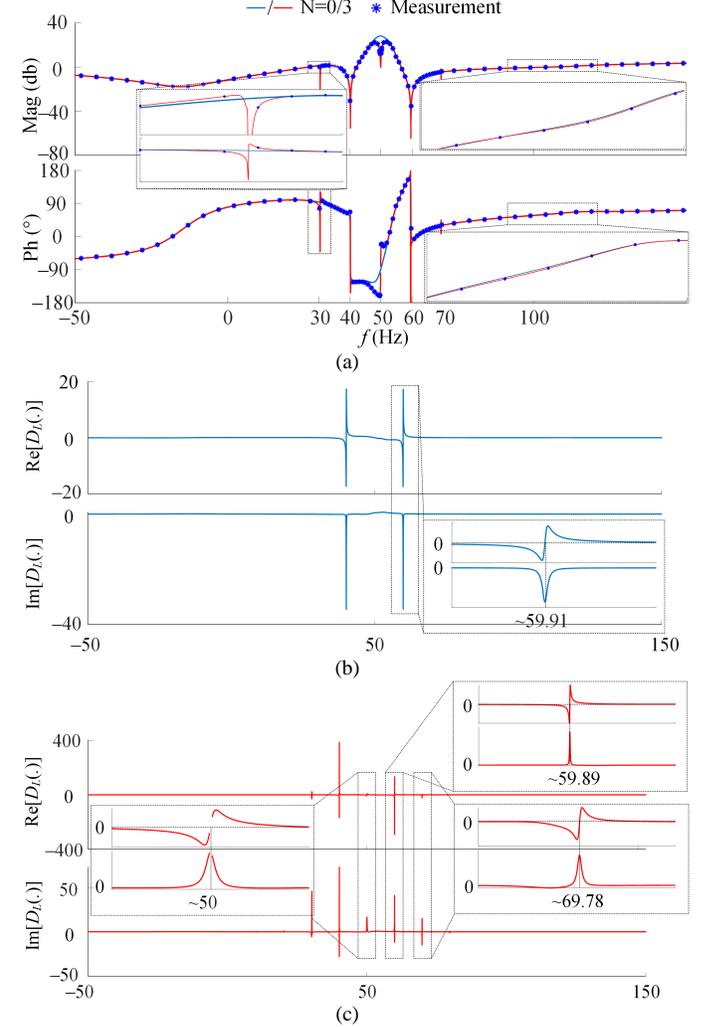

Fig. 9. Linear analysis for Test 1. (a) Frequency responses of $Z_{loop}(s)$. (b)-(c) Mode identification (N=0/3).

In Fig. 9 (a), the red solid line (theoretical derivation with N=3) matches well with the blue asterisks (single-tune frequency scan) over the range of −50-150 Hz, which proves the correctness of the proposed model. The blue solid line (theoretical derivation with N=0) is confirmed to be equal to the existing impedance models [6]-[9]. Obvious discrepancies exist for both the amplitude and phase responses between the two solid lines over the range of 40-60 Hz. Such a phenomenon shows the critical effect of steady-state harmonics on the system stability.

The system modes are further identified by calculating $D_L(Z_{loop}(s))$. A pair of negative damping modes is identified in Fig. 9 (b) and explains the divergence in Fig. 4. However, more zeros and poles around $f_1$, ($f_1 \pm f_s$), ($f_1 \pm 2f_s$), ⋯ are observed for SO in Fig. 9 (c) due to the peaks of frequency responses. All the identified modes own positive damping, so the calculated SO exists and can transit from/to another steady state as Fig. 4 (d) shows instead of being critically stable.

### C. Loop Impedance and Mode Identification (Test 4)

As mentioned in Appendix A. B, even if the theoretical $Z_{\text{loop}}(s)$ for the hard limit triggered case is not derived in this work, the effect of the hard limit on the system mode can be studied using Fig. 10 for Test 4. The blue solid line in Fig. 10 (a) follows the same rule as that in Fig. 9 (b) but has obvious errors compared with the frequency scans.

In Fig. 10 (b), a ~60.1 Hz negative damping mode of the analytical $Z_{\text{loop}}(s)$ without the triggered hard limit modeled is replaced by a ~58.8 Hz positive damping mode of the real $Z_{\text{loop}}(s)$. Such a phenomenon explains the convergence of Fig. 1 (b). Therefore, the effects of the hard limit on the small-signal stability include a) altering the steady-state harmonics, as (A6) should be added to the iteration, and b) offering extra damping to the existing modes instead of adding new modes. For Tests 2-4, the hard limits generally offer positive damping since the absolute value of Im[$D_L(\cdot)$] decreases with a narrower interval between $l_{\text{up}}$ and $l_{\text{low}}$. However, there is a possibility that the hard limit offers negative damping and drives the system to diverge [24]. Such an indication reflects the importance of performing the authentic linear analysis on the hard limit case in the future. The basic idea is following the similar input-output modeling principle in Section II. C and adjusting $E(s)$ in (B4).

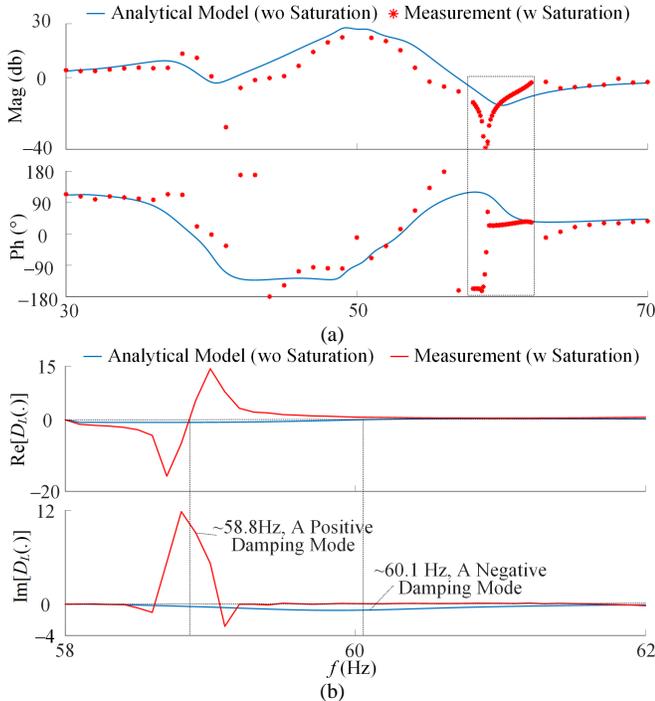

Fig. 10. Loop impedance and mode identification for Test 4. (a) Frequency responses of $Z_{\text{loop}}(s)$. (b) Mode identification.

## VI. DISCUSSION AND CONCLUSION

### A. Comparison among Various-Type SO Analyses

To faithfully reveal the pros and cons of the proposed methods based on harmonic balance compared to the DF-based method with [21]-[23] (or without [15], [17]-[19], [20]) Newton-Raphson iteration embedded, three aspects including accuracy, scalability, complexity are listed in Table IV. Annotations are added below:

a) Since various-type nonlinearities and the high-order components of the complete operating trajectory are considered, the proposed method should have the highest accuracy among the three methods. As the results in Appendix C show, the approximations on nonlinearities have minimal impact on the calculation accuracy.

b) Improving computational accuracy comes at the cost of increasing algorithm complexity and missing the convenience of the graphical prediction of the non-iteration-based method. The proposed method should serve as the reference of the early proposed SO analysis since harmonic balance is also the theoretical basis of DF. A meaningful improvement for this work is to learn from some mature algorithms (such as fast decoupled power flow and dc power flow [28]) and to integrate the graphical prediction. A quantitative comparison between DF-based methods and the harmonic balance-based method is also valuable.

c) The high scalability of the proposed method can be understood from many perspectives. Since the harmonic balance holds for any steady state of the nonlinear time-periodic (NTP) system, the establishment of proposed methods should be regardless of converter topologies. Since a systematic treatment on the triggered hard limit is offered in Section III. C, even if only a single d-axis outer loop control is considered, the method can be extended to a multiple hard limit triggered case using a reliable identification and adding extra variables. Hence, one of the major contributions of this work is successfully solving the hard limit non-triggered case. The authentic linear analysis on SO using the accurate steady-state harmonics also offers a typical example of the extended multiharmonic linearization theory, since the multifrequency coupling is becoming an intrinsic property for a converter-based generation system.

TABLE IV
COMPARISON AMONG VARIOUS-TYPE SO ANALYSES

|  | DF-based (wo iteration) | DF-based (iteration) | Harmonic balance-based (Proposed) |
|---|---|---|---|
| Accuracy | Low | Medium | High |
| Complexity | Low | Medium | High |
| Scalability | Medium | Medium | High |
| Comments | Graphical prediction | -- | For linear analysis |

### B. Insights of the Work

Fig. 11 illustrates the twofold insights of this work:

a) Based on the comparison between various $D_L(Z_{\text{loop}})$s in Figs. 9 and 10, Fig. 11 (a) summarizes the development of the initial negative damping mode to the final positive damping mode. It can offer some new perspectives to oscillation mitigation. If one regards the steady state with/without harmonics flow in the system as "SO on/off", the linear analysis should be applied for both "SO on" and "SO off" since it is important to avoid the divergence at the moment of the mitigation put into operation. The practical mitigation is recommended to be configured between the "SO on" and "SO off" since the mitigation tends to be applied before the system enters SO. Moreover, for the dedicated shunt-connected compensation device, the effect of the steady-state harmonic current with its contributed dynamics should be reflected in the linear analysis for "SO off".

b) Based on the investigation of SO, Fig. 11 (b) describes the relationship between steady-state calculation and linear analysis for an arbitrary NTP system. Since most of the reported works focus on Stage I, Stage II is mostly addressed in this work while extensions are expected to be given on Stage III if necessary. Apart from the cognition that the steady state serves as the basis for linear analysis, exploiting the information of linear analysis will accelerate Newton-Raphson iteration (as Section III. B shows) and determine whether starting a new stage of analysis.

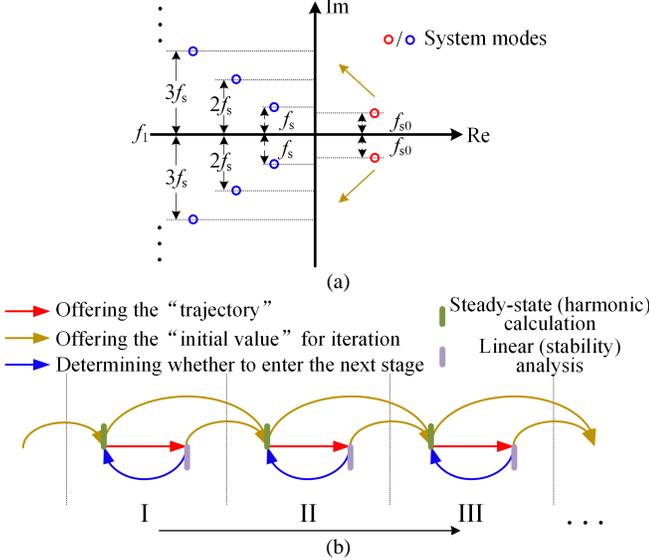

Fig. 11. Insights of this work. (a) The development from negative to positive damping modes for a stable SO. (b) The relationship between steady-state calculation and linear (stability) analysis for an arbitrary NTP system.

## C. Conclusion

Regarding the negative damping-induced SO, existing analytical methods are mostly used for an incomplete steady-state calculation without proof of its existence. This work aims to offer new perspectives to the research field with feasible approaches. Critical observations are initially given to unveil overlooked details of SO. Through a comprehensive analysis of the control and power stages of converters, a set of nonlinear equations is established considering the typical soft and hard nonlinearities as well as higher-order harmonics. The equations are then effectively solved using Newton-Raphson iteration with appropriately determined initial values. To further understand SO, an extended multiharmonic linearization is applied to the loop impedance. The analysis results in the identification of a series of positive damping modes, contrasting with the commonly observed negative or zero damping modes associated with SO. Simulation results effectively validate the rationality of the proposed methods. Furthermore, insights of this work are discussed and future works are also prospected.

## APPENDIX

### A. Details of Nonlinear Equation Establishment

#### a) Trigonometric function approximation

$\theta(t)$ can be expressed using the components of $\boldsymbol{\theta}_0$:

$$\theta(t)=2\pi f_1 t+\theta_0(t)=\omega_1 t+\mathrm{M}(\boldsymbol{\theta}_0\langle 0\rangle)+2\sum_{n=1}^{N}\mathrm{M}(\boldsymbol{\theta}_0\langle \mathrm{ns}\rangle)\cos[\mathrm{n}\omega_1 t+\mathrm{A}(\boldsymbol{\theta}_0\langle \mathrm{ns}\rangle)]$$

(A1)

The Jacobi-Anger expansion of trigonometric functions is:

$$\cos\{\mathrm{M}(\cdot)\cos[\mathrm{A}(\cdot)]\}=J_0[\mathrm{M}(\cdot)]+2\sum_{n=1}^{N}(-1)^n J_{2n}[\mathrm{M}(\cdot)]\cos[2\mathrm{n}\mathrm{A}(\cdot)]$$

$$\sin\{\mathrm{M}(\cdot)\cos[\mathrm{A}(\cdot)]\}=-2\sum_{n=1}^{N}(-1)^n J_{2n-1}[\mathrm{M}(\cdot)]\cos[(2\mathrm{n}-1)\mathrm{A}(\cdot)]$$

(A2)

where $J_n(\cdot)$ is the $\mathrm{n^{th}}$-order Bessel function [24]. Considering the theoretical maximum of $\mathrm{M}(\theta_0(\mathrm{s}))$ ($\sim 0.25\pi$, proved by the active power limit), the tendency of exponentially decrease for the amplitude of high-order components, and the schematic diagram of the Bessel function shown in Fig. 12, it is rational to consider that when $N\geq 3$, the corresponding $J_0(\cdot)$ approaches 1 while $J_n(\cdot)$ ($n\neq 0$) approaches 0. Suppose that N=2, $\cos[\theta(t)]$ and $\sin[\theta(t)]$ are approximated as:

$$\begin{cases}\cos[\theta(t)]=c_1(t)c_s(t)c_{2s}(t)-c_1(t)s_s(t)s_{2s}(t)-s_1(t)s_s(t)c_{2s}(t)-s_1(t)c_s(t)s_{2s}(t)\\ \sin[\theta(t)]=s_1(t)c_s(t)c_{2s}(t)-s_1(t)s_s(t)s_{2s}(t)+c_1(t)s_s(t)c_{2s}(t)+c_1(t)c_s(t)s_{2s}(t)\end{cases},$$

$$\begin{cases}c_1(t)=\cos[\omega_1 t+\mathrm{A}(\boldsymbol{\theta}_0\langle 0\rangle)],\ \ s_1(t)=\sin[\omega_1 t+\mathrm{A}(\boldsymbol{\theta}_0\langle 0\rangle)]\\ c_s(t)\approx J_0[\mathrm{M}(\boldsymbol{\theta}_0\langle \mathrm{s}\rangle)]-2J_2[\mathrm{M}(\boldsymbol{\theta}_0\langle \mathrm{s}\rangle)]\cos[2\omega_1 t+\mathrm{A}(\boldsymbol{\theta}_0\langle \mathrm{s}\rangle)]\\ c_{2s}(t)\approx J_0[\mathrm{M}(\boldsymbol{\theta}_0\langle 2\mathrm{s}\rangle)]-2J_2[\mathrm{M}(\boldsymbol{\theta}_0\langle 2\mathrm{s}\rangle)]\cos[2\{2\omega_1 t+\mathrm{A}(\boldsymbol{\theta}_0\langle 2\mathrm{s}\rangle)]\}\\ s_s(t)\approx 2J_1[\mathrm{M}(\boldsymbol{\theta}_0\langle \mathrm{s}\rangle)]\cos[\omega_1 t+\mathrm{A}(\boldsymbol{\theta}_0\langle \mathrm{s}\rangle)]\\ s_{2s}(t)\approx 2J_1[\mathrm{M}(\boldsymbol{\theta}_0\langle 2\mathrm{s}\rangle)]\cos[2\omega_1 t+\mathrm{A}(\boldsymbol{\theta}_0\langle 2\mathrm{s}\rangle)]\end{cases}$$

(A3)

Expressing $\boldsymbol{\theta}_0$ in polar coordinates as Table II matches the requirement of the Bessel function. Transforming $c(t)/s(t)$ to $\boldsymbol{c}/\boldsymbol{s}$ and then $\boldsymbol{T}(\boldsymbol{c})/\boldsymbol{T}(\boldsymbol{s})$, $\cos(\theta)/\sin(\theta)$ is estimated by:

$$\begin{cases}\cos(\boldsymbol{\theta})\approx \boldsymbol{T}(\boldsymbol{c}_1)\boldsymbol{T}(\boldsymbol{c}_s)\boldsymbol{c}_{2s}-\boldsymbol{T}(\boldsymbol{c}_1)\boldsymbol{T}(\boldsymbol{s}_s)\boldsymbol{s}_{2s}-\boldsymbol{T}(\boldsymbol{s}_1)\boldsymbol{T}(\boldsymbol{s}_s)\boldsymbol{c}_{2s}-\boldsymbol{T}(\boldsymbol{s}_1)\boldsymbol{T}(\boldsymbol{c}_s)\boldsymbol{s}_{2s}\\ \sin(\boldsymbol{\theta})\approx \boldsymbol{T}(\boldsymbol{s}_1)\boldsymbol{T}(\boldsymbol{c}_s)\boldsymbol{c}_{2s}-\boldsymbol{T}(\boldsymbol{s}_1)\boldsymbol{T}(\boldsymbol{s}_s)\boldsymbol{s}_{2s}+\boldsymbol{T}(\boldsymbol{c}_1)\boldsymbol{T}(\boldsymbol{s}_s)\boldsymbol{c}_{2s}+\boldsymbol{T}(\boldsymbol{c}_1)\boldsymbol{T}(\boldsymbol{c}_s)\boldsymbol{s}_{2s}\end{cases}$$

(A4)

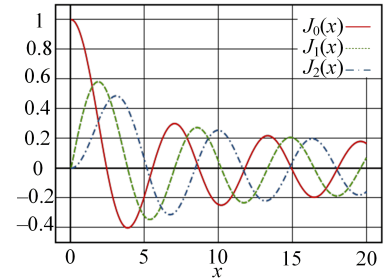

Fig. 12. Schematic diagram of the Bessel function.

#### b) Modeling the triggered hard limit

Here the bilateral triggered case is selected as an example. The time-domain expression of the input to the hard limit in Fig. 3 (b) is expressed as:

$$i_d^{p^*}(t)=\mathrm{M}(i_d^{p^*}\langle 0\rangle)+2\sum_{n=1}^{N}\mathrm{M}(i_d^{p^*}\langle \mathrm{ns}\rangle)\cos[\mathrm{n}\omega_1 t+\mathrm{A}(i_d^{p^*}\langle \mathrm{ns}\rangle)]$$

(A5)

In each harmonic period $T_s$ ($=f_s^{-1}$), the output of the hard limit $i_d(t)$ is divided into 5 parts. Solving the segment moments using the transcendental equations $i_d^{p^*}(t)=l_{up}/l_{low}$ is inefficient. The alternative is to approximate $i_d^{p^*}(t)$ with only the dc and the $\mathrm{1^{st}}$-order harmonic [21], then the segment moments can be expressed using inverse trigonometric functions.

The principle of Fourier series combining the system closed-loop operation ensures the consistency of harmonic frequency distribution between the input and output of the hard limit, as defined in the generalized ZS steady-state vector in (1). Therefore, the harmonics of the output time-domain signals can be obtained via piecewise Fourier series calculation and then transformed into a frequency-domain vector. Such an idea also benefits deriving an analytical expression of DF [21]. By letting $l_{up}/l_{low}=+/-2\,M(\hat{i}_d^{s^*}\langle s\rangle)$, the bilateral triggered case turns into the unilateral or non-triggered case. Hence, the effect of hard limit can be described using the mapping function $\boldsymbol{L}(\cdot)$:

$$\boldsymbol{i}_d^{s^*}=\boldsymbol{L}(\hat{i}_d^{s^*},l_{up},l_{low}) \tag{A6}$$

*c) Expressions for constructing nonlinear equations*

Some intermediate vectors can be derived using the linear relationships contributed by the passive elements and the PI controllers (the parameters in Table I and the vectors $\boldsymbol{u}_{dc}^*$, $\boldsymbol{u}_{ta}$, $\boldsymbol{i}_{load}$, and $\boldsymbol{i}_q^*$ are regarded as known quantities):

$$\begin{cases}\boldsymbol{i}_a=diag(jL_g\boldsymbol{\omega})^{-1}(\boldsymbol{u}_{ga}-\boldsymbol{u}_{ta}) \\ \boldsymbol{i}_{dc}{}'=\boldsymbol{i}_{load}-diag(jC_{dc}\boldsymbol{\omega})\boldsymbol{u}_{dc} \\ \boldsymbol{e}_a^*=\boldsymbol{u}_{ca}+diag(jL_f\boldsymbol{\omega})\boldsymbol{i}_a \\ \boldsymbol{u}_{ga}=\boldsymbol{u}_{ca}\end{cases},\begin{cases}\boldsymbol{e}_d{}'=diag(k_p^{cc}\mathbf{I}+k_i^{cc}./\boldsymbol{\omega})(\boldsymbol{i}_d^*-\boldsymbol{i}_d) \\ \boldsymbol{e}_q{}'=diag(k_p^{cc}\mathbf{I}+k_i^{cc}./\boldsymbol{\omega})(\boldsymbol{i}_q^*-\boldsymbol{i}_q) \\ \boldsymbol{\theta}'=diag(k_p^{pll}\mathbf{I}+k_i^{pll}./\boldsymbol{\omega})\boldsymbol{u}_{gq} \\ \boldsymbol{i}_d^{s*'}=diag(k_p^{dc}\mathbf{I}+k_i^{dc}./\boldsymbol{\omega})(\boldsymbol{u}_{dc}-\boldsymbol{u}_{dc}^*)\end{cases}. \tag{A7}$$

where $\mathbf{I}$ is the identity matrix. "./" represents the right array division in MATLAB. "$^{s}$" indicates an extra expression of the corresponding vector due to the control stage:

$$\begin{cases}\boldsymbol{e}_a=0.5\boldsymbol{T}(\boldsymbol{m}_a)\boldsymbol{u}_{dc} \\ \boldsymbol{i}_{dc}=1.5\boldsymbol{T}(\boldsymbol{m}_a)\boldsymbol{i}_a\end{cases},\begin{cases}\boldsymbol{e}_d=\dfrac{2}{k_{pwm}}\boldsymbol{T}(\boldsymbol{m}_a)\cos(\boldsymbol{\theta}) \\ \boldsymbol{e}_q=\dfrac{2}{k_{pwm}}\boldsymbol{u}_{dc}^*\boldsymbol{T}(\boldsymbol{m}_a)\sin(\boldsymbol{\theta})\end{cases},\begin{cases}\boldsymbol{i}_d=2\boldsymbol{T}(\boldsymbol{i}_a)\cos(\boldsymbol{\theta}) \\ \boldsymbol{i}_q=2\boldsymbol{T}(\boldsymbol{i}_a)\sin(\boldsymbol{\theta}) \\ \boldsymbol{u}_{gq}=2\boldsymbol{T}(\boldsymbol{u}_{ga})\sin(\boldsymbol{\theta})\end{cases}. \tag{A8}$$

Since several dc variables are infinite when substituting $\boldsymbol{\omega}\langle 0\rangle=0$ into (A8), the following relationships must satisfy:

$$\boldsymbol{i}_d^*\langle 0\rangle=\boldsymbol{i}_d\langle 0\rangle,\ \boldsymbol{i}_q^*\langle 0\rangle=\boldsymbol{i}_q\langle 0\rangle. \tag{A9-1}$$

$$\boldsymbol{u}_{gq}\langle 0\rangle=0 \tag{A9-2}$$

$$\boldsymbol{u}_{dc}\langle 0\rangle=\boldsymbol{u}_{dc}^*\langle 0\rangle \tag{A9-3}$$

Moreover, since no active power loss exists in the testing system, the 0 Hz active power balance holds for the ac-dc side of the converter, the PCC, and the ac infinite bus. Therefore, I($\boldsymbol{u}_{ga}\langle 1\rangle$) is a constant for the testing system:

$$\begin{cases}\boldsymbol{u}_{ta}\langle 1\rangle+j\omega_1L_g\boldsymbol{i}_a\langle 1\rangle=\boldsymbol{u}_{ga}\langle 1\rangle \\ I_{load}\boldsymbol{u}_{dc}^*=\mathrm{Re}(3\boldsymbol{i}_a\langle 1\rangle\boldsymbol{i}_{ta}\langle 1\rangle)\end{cases}\Rightarrow I_{load}\boldsymbol{u}_{dc}^*=\mathrm{Re}\left[\dfrac{j3(\boldsymbol{u}_{ta}\langle 1\rangle-\boldsymbol{u}_{ga}\langle 1\rangle)\boldsymbol{u}_{ta}\langle 1\rangle}{\omega_1L_g}\right] \\ \Rightarrow \omega_1L_gI_{load}\boldsymbol{u}_{dc}^*=-j3[\mathrm{I}(\boldsymbol{u}_{ga}\langle 1\rangle)]\Rightarrow \mathrm{I}(\boldsymbol{u}_{ga}\langle 1\rangle)=\dfrac{\omega_1L_gI_{load}\boldsymbol{u}_{dc}^*}{3U_t}. \tag{A10}$$

Such a treatment benefits the low-order and non-sparsity of the Jacobian matrix in Newton-Raphson iteration.

*B. Details of Loop Impedance Derivation*

To extend the theory of multiharmonic linearization, only the hard limit non-triggered case is considered. Multiharmonic linearization on the first group equations of (A7) and (A8) yields the small-signal model of the power stage:

$$\begin{cases}0.5\boldsymbol{T}(\boldsymbol{m}_a^*)\Delta\boldsymbol{u}_{dc}+0.5\boldsymbol{T}(\boldsymbol{v}_{dc}^*)\Delta\boldsymbol{m}_a-\boldsymbol{Z}_f\Delta\boldsymbol{i}_a=\Delta\boldsymbol{u}_{ca} \\ 1.5[\boldsymbol{T}(\boldsymbol{m}_a^*)\Delta\boldsymbol{i}_a+\boldsymbol{T}(\boldsymbol{i}_a^*)\Delta\boldsymbol{m}_a]=\Delta\boldsymbol{i}_{dc}\end{cases} \tag{B1}$$

As $Z_{loop}(s)/Y_{loop}(s)$ describes the relationship of $\Delta\boldsymbol{u}$ and $\Delta\boldsymbol{i}$, the intermediate vectors $\Delta\boldsymbol{m}_a$, $\Delta\boldsymbol{u}_{dc}$, and $\Delta\boldsymbol{u}_{ca}$ are calculated as:

$$\Delta\boldsymbol{m}_a=k_{pwm}[\boldsymbol{P}(s)\Delta\boldsymbol{u}_a+\boldsymbol{Q}(s)\Delta\boldsymbol{i}_a+\boldsymbol{E}(s)\Delta\boldsymbol{u}_{dc}],\ \Delta\boldsymbol{i}_{dc}=-\boldsymbol{Y}_C(s)\Delta\boldsymbol{u}_{dc},$$
$$\Delta\boldsymbol{u}_{ca}=\Delta\boldsymbol{u}_a-\Delta\boldsymbol{u}_{ga}=\Delta\boldsymbol{u}_a-\boldsymbol{Z}_g(s)(-\Delta\boldsymbol{i}_a). \tag{B2}$$

Combining (B1) and (B2), $Y_{loop}(s)$ can be extracted from $\boldsymbol{Y}_{loop}(s)$ with a modular structure:

$$\boldsymbol{Y}_{loop}(s)=-[\mathbf{D}(s)^{-1}\mathbf{C}(s)],\ \boldsymbol{Y}_{loop}(s)=\boldsymbol{Y}_{loop}(s)_{(10N+3,\ 10N+3)} \tag{B3-1}$$

$$\begin{cases}\mathbf{C}(s)=\mathbf{I}-0.5\boldsymbol{T}(\boldsymbol{m}_a^*)-0.5k_{pwm}\boldsymbol{T}(\boldsymbol{u}_{dc}^*)[\boldsymbol{P}(s)+\boldsymbol{E}(s)\boldsymbol{V}(s)]) \\ \mathbf{D}(s)=\boldsymbol{Z}_g(s)+\boldsymbol{Z}_f(s)-0.5\boldsymbol{T}(\boldsymbol{m}_a^*) \\ \quad-0.5k_{pwm}\boldsymbol{T}(\boldsymbol{u}_{dc}^*)\{[\boldsymbol{Q}(s)+\boldsymbol{P}(s)\boldsymbol{Z}_g(s)]+\boldsymbol{E}(s)[\boldsymbol{W}(s)+\boldsymbol{V}(s)\boldsymbol{Z}_g(s)]\}\end{cases} \tag{B3-2}$$

$$\begin{cases}\boldsymbol{U}(s)=-\boldsymbol{Y}_C(s)-1.5k_{pwm}\boldsymbol{T}(\boldsymbol{i}_a^*)\boldsymbol{E}(s) \\ \boldsymbol{V}(s)=1.5k_{pwm}\boldsymbol{U}(s)^{-1}\boldsymbol{T}(\boldsymbol{i}_a^*)\boldsymbol{P}(s) \\ \boldsymbol{W}(s)=1.5\boldsymbol{U}(s)^{-1}[\boldsymbol{T}(\boldsymbol{m}_a^*)+k_{pwm}\boldsymbol{T}(\boldsymbol{i}_a^*)\boldsymbol{Q}(s)]\end{cases} \tag{B3-3}$$

$\boldsymbol{Q}(s)$ and $\boldsymbol{E}(s)$ have the concise forms:

$$\begin{cases}\boldsymbol{E}(s)=\boldsymbol{T}(\cos(\boldsymbol{\theta}^*))\boldsymbol{H}_{cc}(s)\boldsymbol{H}_{dc}(s) \\ \boldsymbol{Q}(s)=2\boldsymbol{T}(\cos(\boldsymbol{\theta}^*))[-\boldsymbol{H}_{cc}(s)]\boldsymbol{T}(\cos(\boldsymbol{\theta}^*))+2\boldsymbol{T}(\sin(\boldsymbol{\theta}^*))[-\boldsymbol{H}_{cc}(s)]\boldsymbol{T}(\sin(\boldsymbol{\theta}^*))\end{cases} \tag{B4}$$

In conventional impedance modeling, the only element of the diagonal matrix $\boldsymbol{T}(\cos(\boldsymbol{\theta}''))/\boldsymbol{T}(\sin(\boldsymbol{\theta}''))$ is $\pm0.5/\pm j0.5$, which makes it possible to obtain explicit expression of $\boldsymbol{Q}(s)$ and $\boldsymbol{E}(s)$ [8], [9]. With a more complicated frequency coupling introduced by the Park transformation of SO, it is important to understand and follow the proposed methods in this work.

### C. Steady-State Harmonics Calculations

**TABLE CII**
STEADY-STATE HARMONICS (TEST 2)

| Variable | Mea (Ref) | Iter (N=3) | Variable | Mea (Ref) | Iter (N=3) |
|---|---|---|---|---|---|
| I($m_s(1)$) | 0.0698 | 0.0698 | R($m_s(1)$) | 0.4062 | 0.4063 |
| I($m_s(1-s)$) | -0.0104 | -0.0105 | R($m_s(1-s)$) | -0.0146 | -0.0148 |
| I($m_s(1-2s)$) | 0.0001 | 0.0001 | R($m_s(1-2s)$) | 0.0005 | 0.0006 |
| I($m_s(1-3s)$) | -0.0000 | -0.0000 | R($m_s(1-3s)$) | -0.0000 | -0.0000 |
| I($m_s(1+s)$) | -0.0055 | -0.0056 | R($m_s(1+s)$) | 0.0005 | 0.0005 |
| I($m_s(1+2s)$) | -0.0005 | -0.0005 | R($m_s(1+2s)$) | -0.0004 | -0.0004 |
| I($m_s(1+3s)$) | 0.0000 | 0.0000 | R($m_s(1+3s)$) | -0.0000 | -0.0000 |
| I($u_{dc}(s)$) | -19.2614 | -19.5120 | R($u_{dc}(s)$) | 21.9410 | 22.2239 |
| I($u_{dc}(2s)$) | -0.4490 | -0.4728 | R($u_{dc}(2s)$) | -0.4827 | -0.5102 |
| I($u_{dc}(3s)$) | 0.0044 | -0.0158 | R($u_{dc}(3s)$) | -0.0128 | 0.0051 |
| I($u_{ga}(1-s)$) | 0.5081 | 0.5146 | R($u_{ga}(1)$) | 153.1952 | 153.2011 |
| I($u_{ga}(1-2s)$) | -0.0948 | -0.1000 | R($u_{ga}(1-s)$) | -1.0973 | -1.1119 |
| I($u_{ga}(1-3s)$) | 0.0002 | -0.0023 | R($u_{ga}(1-2s)$) | 0.0165 | 0.0171 |
| I($u_{ga}(1+s)$) | -3.3640 | -3.4069 | R($u_{ga}(1-3s)$) | 0.0001 | -0.0020 |
| I($u_{ga}(1+2s)$) | -0.2245 | -0.2362 | R($u_{ga}(1+s)$) | 3.4376 | 3.4803 |
| I($u_{ga}(1+3s)$) | 0.0096 | 0.0044 | R($u_{ga}(1+2s)$) | -0.1739 | -0.1839 |
| A($\tilde{i}_r(s)$) | -2.2293 | -2.2293 | R($u_{ga}(1+3s)$) | -0.0219 | -0.0176 |
| A($\tilde{i}_r(2s)$) | 2.4434 | 2.4417 | A($\theta_0(s)$) | 3.3279 | 3.3264 |
| A($\tilde{i}_r(3s)$) | 1.4270 | 0.4964 | M($\theta_0(0)$) | 0.1095 | 0.1095 |
| M($\tilde{i}_r(0)$) | 108.1386 | 108.1452 | M($\theta_0(s)$) | 0.0171 | 0.0173 |
| M($\tilde{i}_r(s)$) | 47.0950 | 47.7159 | M($\theta_0(2s)$) | 0.0010 | 0.0010 |
| M($\tilde{i}_r(2s)$) | 0.5348 | 0.5644 | $f_s$ | 9.8750 | 9.8833 |
| M($\tilde{i}_r(3s)$) | 0.0074 | 0.0091 | | | |

**TABLE CIII**
STEADY-STATE HARMONICS (TEST 3)

| Variable | Mea (Ref) | Iter (N=3) | Variable | Mea (Ref) | Iter (N=3) |
|---|---|---|---|---|---|
| I($m_s(1)$) | 0.0924 | 0.0924 | R($m_s(1)$) | 0.3981 | 0.3981 |
| I($m_s(1-s)$) | -0.0060 | -0.0060 | R($m_s(1-s)$) | -0.0306 | -0.0306 |
| I($m_s(1-2s)$) | 0.0001 | 0.0001 | R($m_s(1-2s)$) | 0.0004 | 0.0004 |
| I($m_s(1-3s)$) | -0.0001 | 0.0004 | R($m_s(1-3s)$) | -0.0004 | -0.0006 |
| I($m_s(1+s)$) | -0.0039 | -0.0040 | R($m_s(1+s)$) | 0.0099 | 0.0099 |
| I($m_s(1+2s)$) | -0.0035 | -0.0036 | R($m_s(1+2s)$) | 0.0048 | 0.0048 |
| I($m_s(1+3s)$) | -0.0015 | -0.0010 | R($m_s(1+3s)$) | -0.0003 | -0.0003 |
| I($u_{dc}(s)$) | -9.1153 | -9.1206 | R($u_{dc}(s)$) | 47.2927 | 47.4322 |
| I($u_{dc}(2s)$) | -1.1964 | -1.2400 | R($u_{dc}(2s)$) | 2.2949 | 2.3215 |
| I($u_{dc}(3s)$) | -0.3418 | 0.1168 | R($u_{dc}(3s)$) | 0.2819 | 0.4105 |
| I($u_{ga}(1-s)$) | 1.2887 | 1.2964 | R($u_{ga}(1)$) | 150.4961 | 150.4972 |
| I($u_{ga}(1-2s)$) | 0.0819 | 0.0964 | R($u_{ga}(1-s)$) | -1.7858 | -1.7898 |
| I($u_{ga}(1-3s)$) | 0.0261 | 0.1023 | R($u_{ga}(1-2s)$) | -0.1113 | -0.1165 |
| I($u_{ga}(1+s)$) | -0.8539 | -0.8542 | R($u_{ga}(1-3s)$) | -0.0899 | -0.1284 |
| I($u_{ga}(1+2s)$) | -0.1899 | -0.1246 | R($u_{ga}(1+s)$) | 9.9752 | 10.0060 |
| I($u_{ga}(1+3s)$) | -0.5271 | -0.3336 | R($u_{ga}(1+2s)$) | 1.8356 | 1.8561 |
| A($\tilde{i}_r(s)$) | -1.7056 | -1.7052 | R($u_{ga}(1+3s)$) | 0.0515 | 0.0593 |
| A($\tilde{i}_r(2s)$) | -1.9406 | -1.9506 | A($\theta_0(2s)$) | -2.3852 | -2.3910 |
| A($\tilde{i}_r(3s)$) | -2.2865 | -1.1283 | M($\theta_0(0)$) | 0.1659 | 0.1659 |
| M($\tilde{i}_r(0)$) | 135.6834 | 135.6111 | M($\theta_0(s)$) | 0.0379 | 0.038 |
| M($\tilde{i}_r(s)$) | 86.7301 | 86.9757 | M($\theta_0(2s)$) | 0.0062 | 0.0063 |
| M($\tilde{i}_r(2s)$) | 2.3409 | 2.3806 | $f_s$ | 8.8500 | 8.8522 |
| M($\tilde{i}_r(3s)$) | 0.2692 | 0.2593 | | | |

**TABLE CIV**
STEADY-STATE HARMONICS (TEST 4)

| Variable | Mea (Ref) | Iter (N=3) | Variable | Mea (Ref) | Iter (N=3) |
|---|---|---|---|---|---|
| I($m_s(1)$) | 0.0924 | 0.0924 | R($m_s(1)$) | 0.3981 | 0.3981 |
| I($m_s(1-s)$) | -0.0273 | -0.0275 | R($m_s(1-s)$) | -0.0087 | -0.0088 |
| I($m_s(1-2s)$) | 0.0010 | 0.0010 | R($m_s(1-2s)$) | 0.0000 | -0.0000 |
| I($m_s(1-3s)$) | 0.0001 | -0.0004 | R($m_s(1-3s)$) | 0.0007 | 0.0013 |
| I($m_s(1+s)$) | 0.1659 | 0.1659 | R($m_s(1+s)$) | 0.0012 | 0.0012 |
| I($m_s(1+2s)$) | -0.0008 | -0.0007 | R($m_s(1+2s)$) | -0.0032 | -0.0033 |
| I($m_s(1+3s)$) | 0.0018 | 0.0013 | R($m_s(1+3s)$) | 0.0001 | 0.0009 |
| I($u_{dc}(s)$) | -42.4578 | -42.6686 | R($u_{dc}(s)$) | 13.5175 | 13.6314 |
| I($u_{dc}(2s)$) | -0.1083 | -0.0866 | R($u_{dc}(2s)$) | -1.6015 | -1.6478 |
| I($u_{dc}(3s)$) | 0.4546 | -0.0151 | R($u_{dc}(3s)$) | -0.3035 | -0.6512 |
| I($u_{ga}(1-s)$) | -0.8489 | -0.8481 | R($u_{ga}(1)$) | 150.5553 | 150.5558 |
| I($u_{ga}(1-2s)$) | -0.0532 | -0.0670 | R($u_{ga}(1-s)$) | -1.8337 | -1.8488 |
| I($u_{ga}(1-3s)$) | -0.0323 | -0.0959 | R($u_{ga}(1-2s)$) | 0.1447 | 0.1405 |
| I($u_{ga}(1+s)$) | -8.5376 | -8.5952 | R($u_{ga}(1-3s)$) | 0.1582 | 0.2397 |
| I($u_{ga}(1+2s)$) | -0.3192 | -0.3120 | R($u_{ga}(1+s)$) | 3.7156 | 3.7420 |
| I($u_{ga}(1+3s)$) | 0.6127 | 0.3730 | R($u_{ga}(1+2s)$) | -1.2688 | -1.3065 |
| A($\tilde{i}_r(s)$) | -2.7776 | -2.7766 | R($u_{ga}(1+3s)$) | 0.2065 | 0.1347 |
| A($\tilde{i}_r(2s)$) | 1.7345 | 1.7494 | A($\theta_0(s)$) | 1.4990 | 1.4912 |
| A($\tilde{i}_r(3s)$) | 0.7599 | 0.7544 | M($\theta_0(0)$) | 0.1659 | 0.1659 |
| M($\tilde{i}_r(0)$) | 121.1721 | 121.1716 | M($\theta_0(s)$) | 0.0351 | 0.0353 |
| M($\tilde{i}_r(s)$) | 80.0074 | 80.3491 | M($\theta_0(2s)$) | 0.0039 | 0.0040 |
| M($\tilde{i}_r(2s)$) | 1.4478 | 1.4868 | $f_s$ | 8.8750 | 8.8864 |
| M($\tilde{i}_r(3s)$) | 0.3312 | 0.3943 | | | |

TABLE CI
STEADY-STATE HARMONICS (TEST 1)

| Variable | Mea (Ref) | Iter (N=3) | Iter (N=2) | Iter (N=1) | Iter (N=0) | Variable | Mea (Ref) | Iter (N=3) | Iter (N=2) | Iter (N=1) | Iter (N=0) |
|---|---|---|---|---|---|---|---|---|---|---|---|
| $I(m_a(1))$ | 0.0700 | 0.0700 | 0.0700 | 0.0699 | 0.0698 | $R(m_a(1))$ | 0.4069 | 0.4069 | 0.4069 | 0.4064 | 0.4060 |
| $I(m_a(1-s))$ | 0.0264 | 0.0264 | 0.0263 | 0.0185 | -- | $R(m_a(1-s))$ | 0.0223 | 0.0222 | 0.0222 | 0.0156 | -- |
| $I(m_a(1-2s))$ | 0.0013 | 0.0013 | 0.0013 | -- | -- | $R(m_a(1-2s))$ | 0.0018 | 0.0018 | 0.0017 | -- | -- |
| $I(m_a(1-3s))$ | 0.0000 | 0.0002 | -- | -- | -- | $R(m_a(1-3s))$ | 0.0002 | 0.0001 | -- | -- | -- |
| $I(m_a(1+s))$ | 0.0105 | 0.0104 | 0.0104 | 0.0074 | -- | $R(m_a(1+s))$ | 0.0017 | 0.0017 | 0.0017 | 0.0012 | 0.0012 |
| $I(m_a(1+2s))$ | −0.0006 | −0.0006 | −0.0006 | -- | -- | $R(m_a(1+2s))$ | −0.0020 | −0.0020 | −0.0020 | -- | -- |
| $I(m_a(1+3s))$ | −0.0002 | −0.0001 | -- | -- | -- | $R(m_a(1+3s))$ | −0.0000 | 0.0000 | -- | -- | -- |
| $I(u_{dc}(s))$ | 46.3599 | 46.2193 | 46.1786 | 32.5638 | -- | $R(u_{dc}(s))$ | −31.4876 | −31.3859 | −31.3578 | −21.9784 | -- |
| $I(u_{dc}(2s))$ | −0.4680 | −0.4561 | −0.4498 | -- | -- | $R(u_{dc}(2s))$ | −2.5048 | −2.4973 | −2.5263 | -- | -- |
| $I(u_{dc}(3s))$ | −0.1336 | −0.0270 | -- | -- | -- | $R(u_{dc}(3s))$ | −0.0912 | −0.0369 | -- | -- | -- |
| $I(u_{ga}(1))$ | −0.3960 | −0.3949 | −0.3942 | −0.2867 | -- | $R(u_{ga}(1))$ | 152.9957 | 153.0042 | 153.0047 | 153.1435 | 153.2764 |
| $I(u_{ga}(1-2s))$ | −0.2893 | −0.2902 | −0.2905 | -- | -- | $R(u_{ga}(1-s))$ | 2.2745 | 2.2684 | 2.2664 | 1.6151 | -- |
| $I(u_{ga}(1-s))$ | −0.0288 | −0.0177 | -- | -- | -- | $R(u_{ga}(1-2s))$ | 0.2497 | 0.2481 | 0.2510 | -- | -- |
| $I(u_{ga}(1+s))$ | 7.9200 | 7.8933 | 7.8858 | 5.5617 | -- | $R(u_{ga}(1-3s))$ | 0.0140 | 0.0097 | -- | -- | -- |
| $I(u_{ga}(1+2s))$ | −0.3466 | −0.3415 | −0.3399 | -- | -- | $R(u_{ga}(1+s))$ | −4.7677 | −4.7495 | −4.7448 | −3.3183 | -- |
| $I(u_{ga}(1+3s))$ | −0.0925 | −0.0572 | -- | -- | -- | $R(u_{ga}(1+2s))$ | −0.9802 | −0.9774 | −0.9817 | -- | -- |
| $M(\theta_0(0))$ | 0.1090 | 0.1090 | 0.1090 | 0.1093 | 0.1097 | $R(u_{ga}(1+3s))$ | 0.0112 | 0.0356 | -- | -- | -- |
| $M(\theta_0(s))$ | 0.0328 | 0.0327 | 0.0327 | 0.0230 | -- | $A(\theta_0(2s))$ | 1.7632 | 1.7606 | 1.7579 | -- | -- |
| $M(\theta_0(2s))$ | 0.0036 | 0.0036 | 0.0036 | -- | -- | $f_s$ | 9.9000 | 9.8926 | 9.8923 | 9.9106 | --(9.9132) |

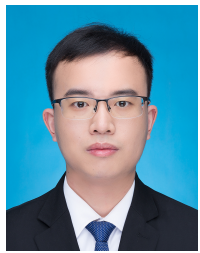

**Chongbin Zhao** (Student Member, IEEE) received the B.S. degree in electrical engineering from Tsinghua University, Beijing, China, in 2019, where he is currently working towards the Ph.D. degree. In 2023, He was a visiting scholar at Rensselaer Polytechnic Institute, Troy, NY, United States. His research interests include power quality analysis and control, and emerging converter-driven power system stability analysis and control.

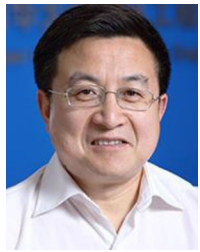

**Qirong Jiang** received the B.S. and Ph.D. degrees in electrical engineering from Tsinghua University, Beijing, China, in 1992 and 1997, respectively. In 1997, he was a Lecturer with the Department of Electrical Engineering, Tsinghua University, where he later became an Associate Professor in 1999. Since 2006, he has been a Professor. His research interests include power system analysis and control, modeling and control of flexible ac transmission systems, power-quality analysis and mitigation, power-electronic equipment, and renewable energy power conversion.